\documentclass[twoside,a4paper,11pt]{sea10}
\usepackage{graphicx}
\usepackage{hyperref}
\usepackage{movie15}
\usepackage{color}
\def\farcs{\hbox{$.\!\!^{\prime\prime}$}}
\def\farcd{\hbox{$.\!\!^{\circ}$}}
\begin{document}
\pagenumbering{arabic}
\pagestyle{myheadings}
\thispagestyle{empty}
{\flushleft\includegraphics[width=\textwidth,bb=58 650 590 680]{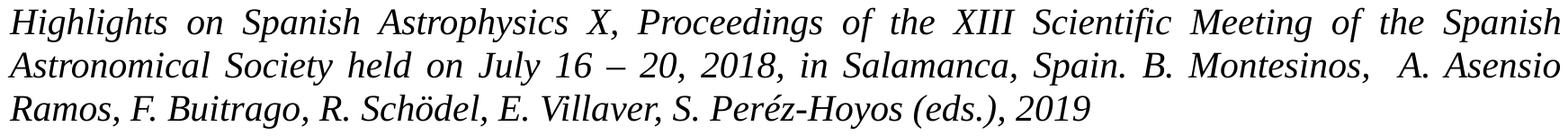}}
\vspace*{0.2cm}
\begin{flushleft}
{\bf {\LARGE
%
%%% TITLE of the paper. 
%%% TITLE of the paper. 
GALANTE: finding all the optically accessible Galactic O+B+WR stars 
in the Galactic Plane
%
% Do not delete next few lines
}\\
\vspace*{1cm}
%
%%% Include here the LIST OF AUTHORS.
%%% Include here the LIST OF AUTHORS.
%%% Note that the last author has to be preceeded by an AND.
J.~Ma{\'\i}z~Apell\'aniz$^1$, 
E.~J.~Alfaro$^{2,3}$, 
R.~H.~Barb\'a$^4$,
A.~Lorenzo$^2$, 
A.~Mar{\'\i}n-Franch$^{3,5}$, 
A.~Ederoclite$^{3,5}$, 
J.~Varela$^{3,5}$, 
H.~V\'azquez~Rami\'o$^{3,5}$, 
J.~Cenarro$^{3,5}$, 
D.~J.~Lennon$^6$, 
and 
P.~Garc{\'\i}a-Lario$^6$
%
% Do not delete next few lines
}\\
\vspace*{0.5cm}
%
%%% AFFILIATIONS LIST.
%%% and the AFFILIATIONS LIST. Note that one affiliation per line.
%%% Add as many affiliations as necessary. 
$^1$ 
Centro de Astrobiolog{\'\i}a, CSIC-INTA, Madrid, Spain\\
$^2$ 
Instituto de Astrof{\'\i}sica de Andaluc{\'\i}a, CSIC, Granada, Spain\\
$^3$ 
Unidad Asociada CEFCA-IAA, CSIC, Teruel, Spain\\
$^4$ 
Universidad de La Serena, La Serena, Chile\\
$^5$ 
Centro de Estudios de F{\'\i}sica del Cosmos de Arag\'on, Teruel, Spain\\
$^6$ 
European Space Agency, ESAC, Madrid, Spain\\
%
% Do not delete next few lines
\end{flushleft}
%
% Headings
\markboth{
%%% Type the SHORT version of the paper title.
%%% Type the SHORT version of the paper title.
GALANTE
}{ % Do not delete
%
%%%  First Author \& Second Author   OR   First-author et al. 
%%%  First Author \& Second Author   OR   First-author et al. if the author list 
%%% contains three or more authors.
Ma{\'\i}z Apell\'aniz et al.
% 
% Do not delete next few lines
}
\thispagestyle{empty}
\vspace*{0.4cm}
\begin{minipage}[l]{0.09\textwidth}
\ 
\end{minipage}
\begin{minipage}[r]{0.9\textwidth}
\vspace{1cm}
\section*{Abstract}{\small
%
% ABSTRACT ABSTRACT ABSTRACT
% ABSTRACT ABSTRACT ABSTRACT
%%% Type the ABSTRACT of your paper
GALANTE is an optical photometric survey with seven intermediate/narrow filters that has been covering the Galactic Plane 
since 2016 using the Javalambre T80 and Cerro Tololo T80S telescopes. The P.I.s of the northern part (GALANTE NORTE) are
Emilio J. Alfaro \& Jes\'us Ma{\'\i}z Apell\'aniz. and the P.I. of the southern part (GALANTE SUR) is Rodolfo H. Barb\'a.
The detector has a continuous 1\farcd4$\times$1\farcd4 field
of view with a sampling of 0\farcs55/pixel and the seven filters are optimized to detect obscured early-type stars.
The survey includes long, intermediate, short, and ultrashort exposure times to reach a dynamical range close to 20 magnitudes, something
never achieved for such an optical project before. The characteristics of GALANTE allow for a new type of calibration scheme
using external Gaia, Tycho-2, and 2MASS data that has already led to a reanalysis of the sensitivity of the Gaia $G$ filter. We
describe the project and present some early results. GALANTE will identify the majority of the early-type massive stars within several
kpc of the Sun and measure their amount and type of extinction. It will also map the H$\alpha$ nebular emission, identify
emission-line stars, and do other studies of low- and intermediate-mass stars.
%
% Do not delete next few lines
\normalsize}
\end{minipage}
%
%
%%% BODY of the paper
%%% BODY of the paper
%
\section{Motivation}

\begin{figure}
%\centerline{\includegraphics[width=\linewidth]{footprint_north.pdf}}
%\centerline{\includegraphics[width=\linewidth]{Maíz_ApellánizJ_3F1a.pdf}}
\centerline{\includegraphics[width=\linewidth]{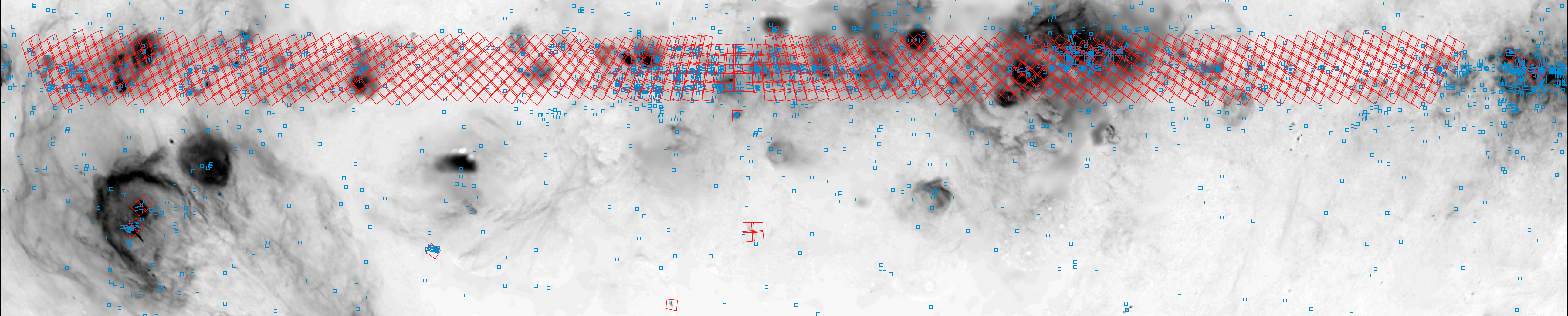}}
\vspace{2mm}
%\centerline{\includegraphics[width=\linewidth]{footprint_south.pdf}}
%\centerline{\includegraphics[width=\linewidth]{Maíz_ApellánizJ_3F1b.pdf}}
\centerline{\includegraphics[width=\linewidth]{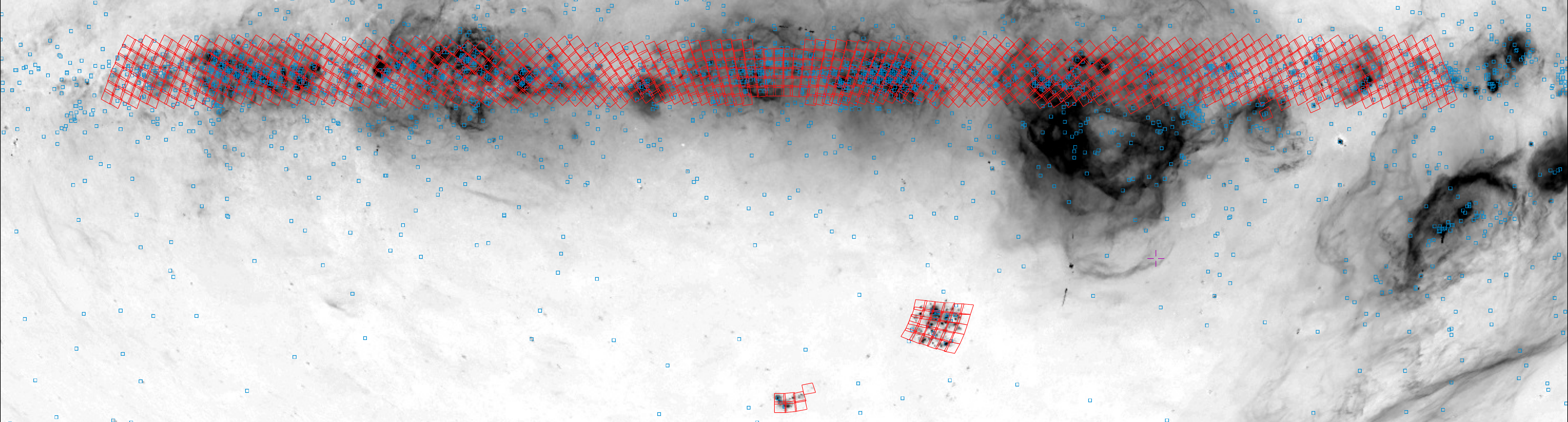}}
\caption{\label{fig1}Footprint (red) of the 2009 GALANTE fields divided by hemisphere (top: north, bottom:south). Blue symbols
         are stars from the Galactic O-Star Catalog (GOSC, \cite{Maizetal04b,Maizetal12,Maizetal17c,Sotaetal08}, 
         \url{http://gosc.cab.inta-csic.es}), which are mostly O+B+WR stars. The background is an H$\alpha$ image
         \cite{Fink03} in a log scale aligned with Galactic coordinates using an Aitoff projection. The off-plane fields 
         include the LMC, the SMC (bottom), M31, and M33 (top).}
\end{figure}

\begin{figure}
\begin{minipage}{0.67\textwidth}
%\centerline{\includegraphics[width=\linewidth]{Pleiades_campaign_001_res.pdf}}
%\centerline{\includegraphics[width=\linewidth]{Maíz_ApellánizJ_3F2.pdf}}
\centerline{\includegraphics[width=\linewidth]{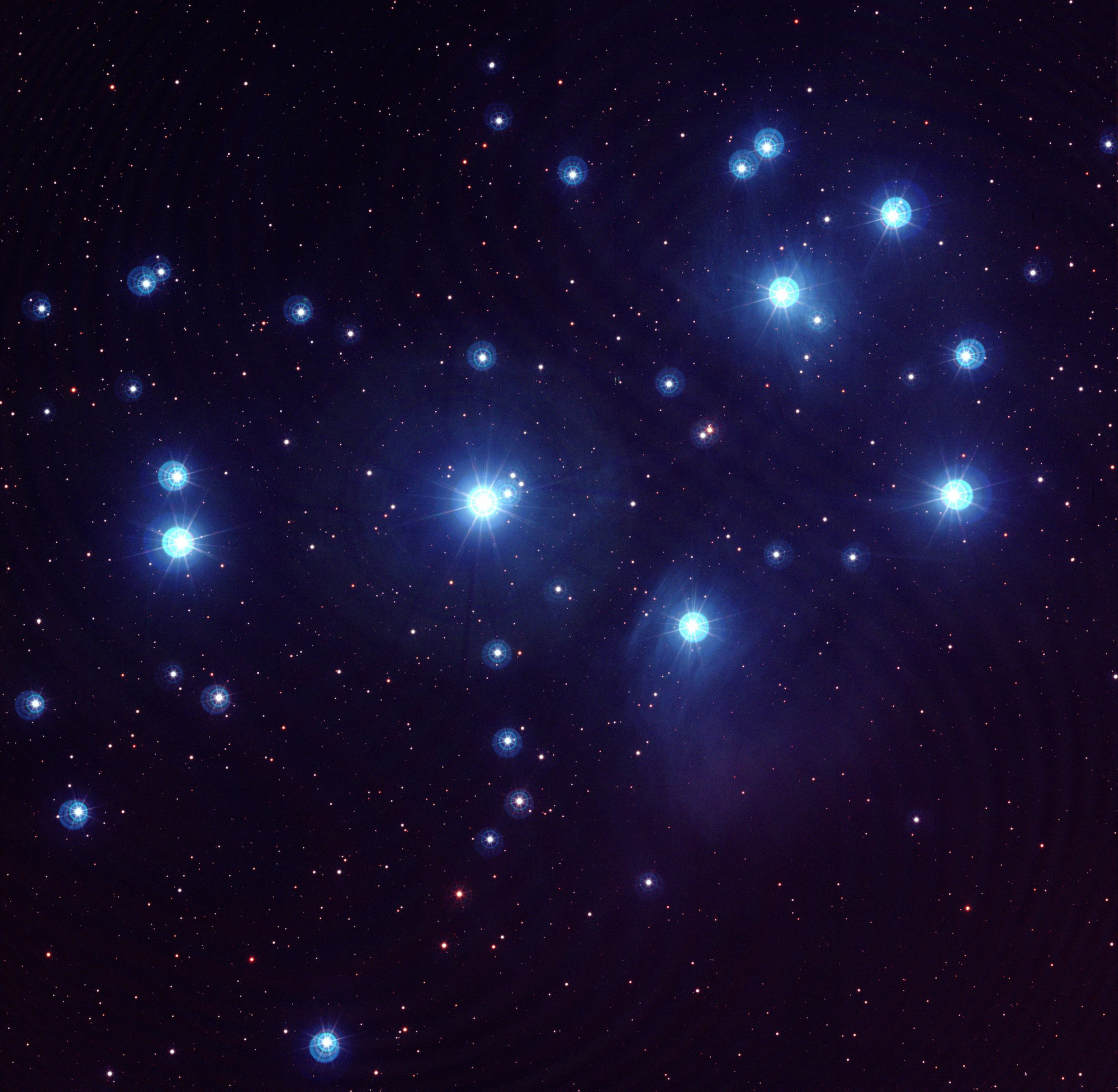}}
\end{minipage}
\begin{minipage}{0.33\textwidth}
\caption{\label{fig2}GALANTE RGB (F665N+F450N+F420N) image of the Pleiades obtained with the Javalambre T80 telescope. The combination of four exposure times 
         (from 0.1~s to 50~s) yields not a single saturated pixel (despite the presence of the 2$^{\rm nd}$ magnitude Alcyone) while detecting 
         faint objects with magnitudes 19-20. The field of view is 1\farcd4$\times$1\farcd4 with N toward the top and E toward the left.}
\end{minipage}
\end{figure}

$\,\!$\indent GALANTE is a project that is imaging the Galactic Plane (Fig.~\ref{fig1}) using the Javalambre T80 and Cerro Tololo T80S twin telescopes 
\cite{Cenaetal17,Maiz17b}. The detector, footprint, exposure times, magnitude range, survey dates, and filters are given in Table~\ref{tab1}. The main goal 
of the project is to identify a near complete sample of Galactic obscured O+B+WR stars up to several kpc and measure their extinction but other 
objectives will be achieved with the survey. As there are a number of other large-area or whole-sky existing/ongoing photometric 
surveys (Tycho-2, 2MASS, SDSS, EGAPS, Gaia, Pan-STARRS, J-PLUS, S-PLUS, J-PAS, VVV\ldots), the reader may ask him/herself: why one more? Because
none has the GALANTE characteristics:

\noindent\textbf{Region of the sky:} Some surveys (e.g. SDSS) concentrate in areas away from the Galactic plane, which they do not cover (or do not do it completely).
GALANTE covers the Galactic Plane, LMC, SMC, M31, M33, and a few off-plane Galactic clusters. \hfill$\,\!$\linebreak
\textbf{Magnitude range:} Most ground-based surveys use 2-4~m telescopes with one or two exposure times. Hence, they saturate around
magnitude 11-12 and cover a dynamic range of 10-12 magnitudes. GALANTE uses 80~cm telescopes with four exposure times to photometer all stars down to 
magnitude 19-20 (Fig.~\ref{fig2}). This is important for obscured stars, which may be close to the detection limit in the blue and be brighter than 
magnitude 10 beyond 8000~\AA. \hfill$\,\!$\linebreak
\textbf{Filter selection:} Some surveys use filter sets not optimized for (e.g. SDSS $ugriz$) or incapable of (e.g. Tycho-2 $BV$ or 2MASS $JHK$) measuring 
the Balmer jump, which is the only realistic method to determine $T_{\rm eff}$ for hot stars \cite{MaizSota08,Maiz13b,Maizetal14a,MaizBarb18}. Also, no similar 
survey includes two narrow filters (on-band and off-band) for H$\alpha$ in order to photometer that line for stars (in absorption or emission) and to map the 
nebulosity with subarcsecond pixels. \hfill$\,\!$\linebreak
\textbf{Calibration accuracy:} GALANTE uses narrow/intermediate band filters, which are less sensitive to photometric accuracy problems induced by atmospheric
extinction. \hfill$\,\!$\linebreak
\textbf{Confusion:} The three-band ($G$+$G_{\rm BP}$+$G_{\rm RP}$) Gaia photometric survey will address most of the problems above, especially after the full
spectrophotometry is available in DR3. However, it will suffer from confusion in crowded and/or nebular regions (common for O+B+WR stars), as the 
$G_{\rm BP}$+$G_{\rm RP}$ instrument behaves as a slitless spectrograph. 

\begin{table}
\centerline{
\begin{tabular}{lll}
{\it Detector:}        & \multicolumn{2}{l}{1\farcd4$\times$1\farcd4 continuous FOV with 0\farcs55 pixels.} \\
{\it Footprint:}       & \multicolumn{2}{l}{T80: $\;\,|l| < 3^{\rm o}$ + $|\delta| > 0^{\rm o}$ plus selected regions, 1100 sq. dg., Fig.~\ref{fig2}.} \\
                       & \multicolumn{2}{l}{T80S: $|l| < 3^{\rm o}$ + $|\delta| < 0^{\rm o}$ plus selected regions, 1100 sq. dg., Fig.~\ref{fig2}.} \\
{\it Exposure times:}  & \multicolumn{2}{l}{2$\times$0.1 s + 2$\times$1 s + 2$\times$10 s + 4$\times$50/100 s (at two different airmasses).} \\
{\it Magnitude range:} & \multicolumn{2}{l}{Unsat. AB mag 3-17 with S/N $>$ 100 in all filters, detect. to AB mag 19-20.} \\
{\it Survey dates:}    & \multicolumn{2}{l}{GALANTE NORTE (T80): 2016-2021. GALANTE SUR (T80S): 2018-2023.} \\
{\it Filters:}         & F348M & Str\"omgren $u$ equivalent, $T_{\rm eff}$ + extinction determination. \\
                       & F420N & Continuum between H$\delta$ and H$\gamma$, $T_{\rm eff}$ + extinction determination. \\
                       & F450N & Continuum between H$\gamma$ and H$\beta$, $T_{\rm eff}$ + extinction determination. \\
                       & F515N & Str\"omgren $y$ equivalent, $T_{\rm eff}$ + extinction determination. \\
                       & F660N & H$\alpha$ line, pure nebular images + emission-line star detection. \\
                       & F665N & H$\alpha$ continuum, pure nebular images + emission-line star detection. \\
                       & F861M & CaT, tie-in with Gaia-RVS and 2MASS, extinction typing. \\
\end{tabular}
}
\caption{GALANTE in a nutshell}
\label{tab1}     
\end{table}

\section{Measuring temperature}

\begin{figure}
\begin{minipage}{0.63\textwidth}
%\centerline{\includegraphics[width=\linewidth]{filters.pdf}}
%\centerline{\includegraphics[width=\linewidth]{Maíz_ApellánizJ_3F3.pdf}}
\centerline{\includegraphics[width=\linewidth]{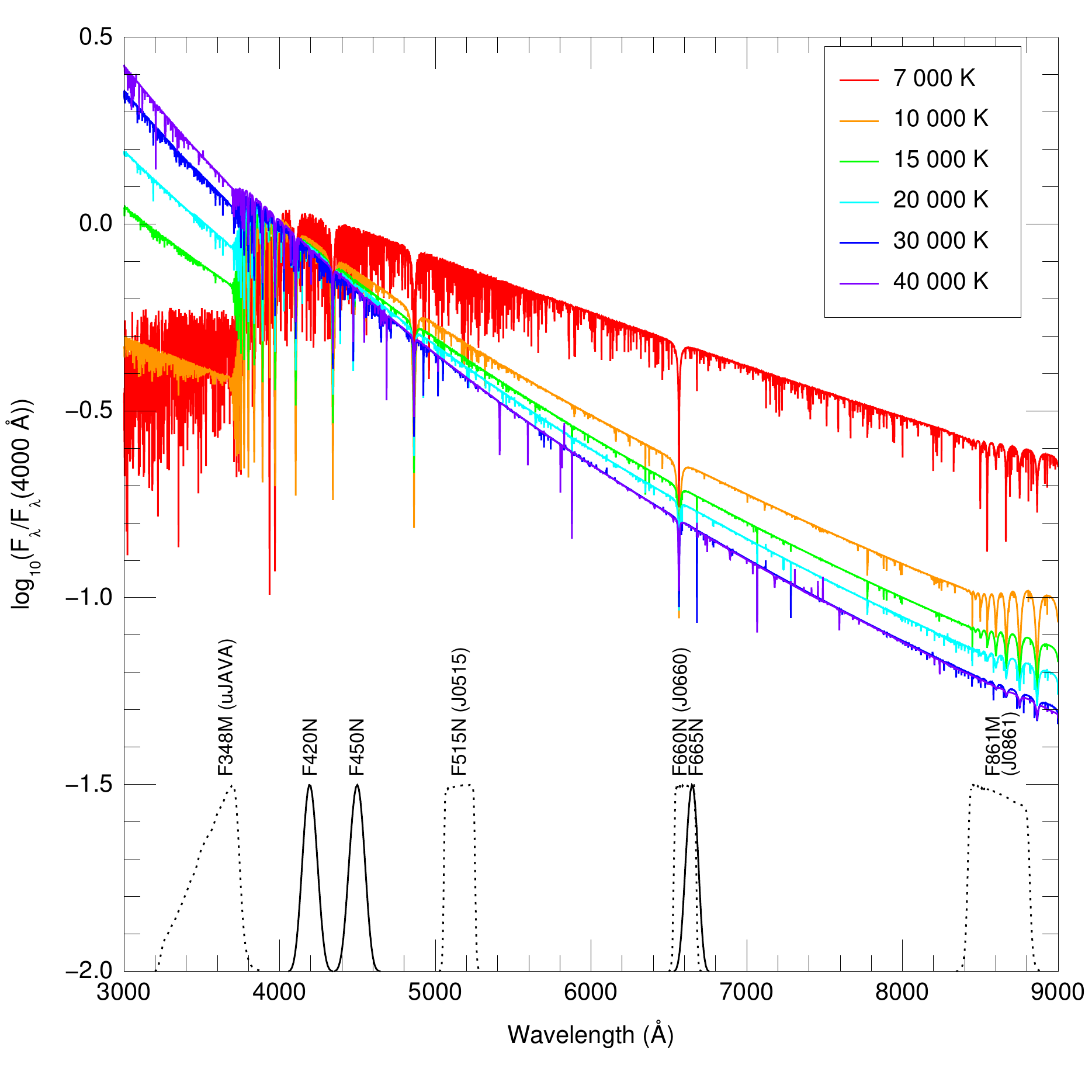}}
\end{minipage}
%\hspace{-0.14\textwidth}
\begin{minipage}{0.35\textwidth}
\caption{\label{fig3}Normalized passbands for the seven GALANTE filters compared with model SEDs of stars with different 
         $T_{\rm eff}$ \cite{Maiz13a}. The filter combination F438M+F420N+F450N+F515N has been chosen to accurately measure the Balmer jump using
         continuum regions that avoid the strong Balmer lines. J-PLUS filters are plotted with dotted lines and filters exclusive to GALANTE are 
         plotted with solid lines.}
\end{minipage}
\end{figure}

$\,\!$\indent Figure~\ref{fig3} shows the seven passbands used in the GALANTE survey, four of them in common with the J-PLUS survey \cite{Cenaetal17}
(F348M+F515N+F660N+F861M) and three specifically designed for the project (F420N+F450N+F665N). The four bluemost filters have been chosen to measure the Balmer
jump: F348M measures the continuum to its left while F420N+F450N+F515N do it to the right avoiding the Balmer lines. GALANTE will be used to measure $T_{\rm eff}$
in two steps.

\begin{figure}
%\centerline{\includegraphics[width=0.49\linewidth]{indexindex.pdf} \
%            \includegraphics[width=0.49\linewidth]{RA_308_73_DEC_+41_62_vis36_4.pdf}}
%\centerline{\includegraphics[width=0.49\linewidth]{Maíz_ApellánizJ_3F4a.pdf} \
%            \includegraphics[width=0.49\linewidth]{Maíz_ApellánizJ_3F4b.pdf}}
\centerline{\includegraphics[width=0.49\linewidth]{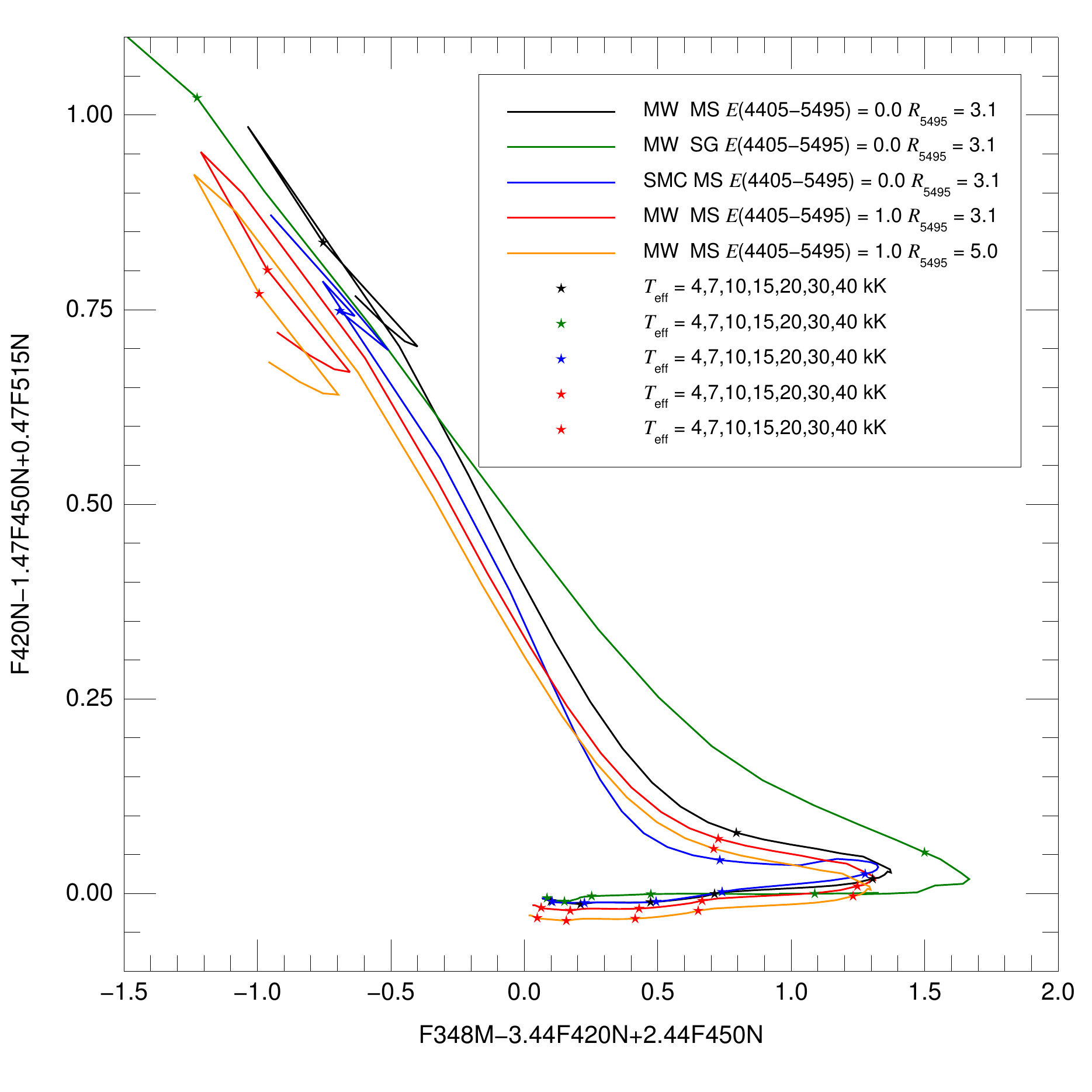} \
            \includegraphics[width=0.49\linewidth]{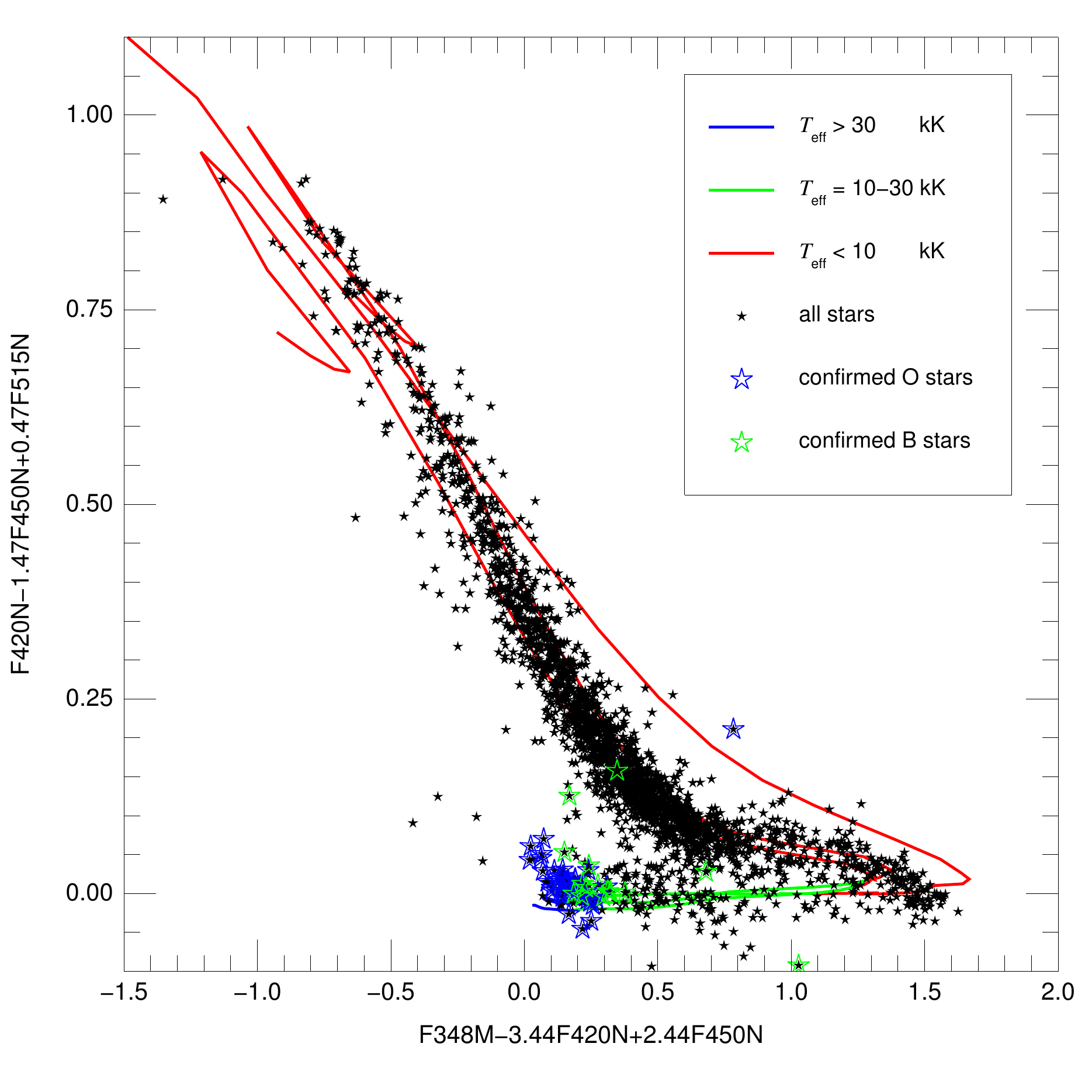}}
\caption{\label{fig4} [left] A index-index diagram using model SEDs from \cite{Maiz13a}. Note how we can measure $T_{\rm eff}$ 
         for hot stars (lower branch) independently of gravity, metallicity, and reddening. [right]
         Same diagram with only the first, second, and fourth synthetic photometry functions plus overplotted data from one of the GALANTE
         fields in Cygnus OB2 (using a preliminary calibration). Most of the stars in the diagram follow the 
         main-sequence track between A and K stars, as expected. The field contains a very rich highly extinguished OB association. 
         This results in few late-B stars present (they are too dim for GALANTE at the distance and extinction of the 
         association) but also in 100+ O and early B stars detected. Despite their high extinction ($A_V \sim 6$~mag), they are at 
         the expected location in the diagram and the existing spectral types confirm their nature, thus providing an indication of 
         the excellent quality of the data and calibration.}
\end{figure}

First, we build the two indices (analogous to $m_1$ and $c_1$ in the Str\"omgren system) $m^\prime$ = F420N$-$1.47\,F450N$+$0.47\,F515N and 
$c^\prime$ = F348M$-$3.44\,F420N$+$2.44\,F450N, where the coefficients were determined empirically. 
As shown in the left panel of Fig.~\ref{fig4}, those indices are nearly independent of gravity, metallicity, and type or amount 
of extinction for hot stars ($T_{\rm eff} > 10$~kK) and depend almost only on $T_{\rm eff}$ with a large dynamic range in $c^\prime$ of more than one magnitude 
between 10~kK and 40~kK. For cooler stars, the position in the index-index diagram depends not only on $T_{\rm eff}$ but also on gravity, and to some extent on 
the other quantities. The validity of this method is shown on the right panel of Fig.~\ref{fig4}, where a preliminary analysis (see below for calibration issues)
of one of the GALANTE fields in Cygnus OB2 using aperture photometry correctly places the stellar locus and classifies the stars with known spectral types. 
Therefore, we will use these indices to give a preliminary estimate of stellar temperature.

As a second step, we will combine GALANTE with 2MASS and process the photometry with CHORIZOS \cite{Maiz04c}, a Bayesian photometric code that 
allows the simultaneous calculation of $T_{\rm eff}$, luminosity class, $E(4405-5495)$ or amount of extinction, and $R_{5495}$ or type of extinction from
photometric data. The output will be combined with Gaia parallaxes to compare trigonometric and spectroscopic distances and build a 3-D extinction map of the solar
vicinity more accurate than previous attempts, allowing for different types of extinction.

\section{Photometric calibration}

$\,\!$\indent The primary photometric calibration of GALANTE uses the fact that a typical 1\farcd4$\times$1\farcd4 Galactic Plane field contains $\sim 100$
objects with good-quality Tycho-2 $B_T$+$V_T$ + Gaia $G$+$G_{\rm BP}$+$G_{\rm RP}$ + 2MASS $J$+$H$+$K$ (8 filters) and $\sim 10^4$ objects with their Gaia and 
2MASS equivalents (6 filters). We process that input photometry with CHORIZOS \cite{Maiz04c} and the SED grid of \cite{Maiz13a} to generate synthetic predicted 
magnitudes (and their uncertainties) in each of the seven GALANTE filters allowing for arbitrary variations in $T_{\rm eff}$, luminosity class, $E(4405-5495)$, 
and $R_{5495}$. The resulting predicted uncertainties for a single star are in the range of one hundredth to a few tenths of magnitude, with lower values for 
objects with 8-filter photometry and lower input uncertainties and for redder filters. For each field and filter we first combine the information from the 
$\sim 10^4$ objects with 6-filter photometry to detect (and correct if necessary) possible low-order flat-field issues and we then use objects with 8-filter
photometry to calculate the zero point, which has a typical uncertainty of one hundredth of a magnitude. The whole process is dependent on the accuracy of the
calibration of the input photometry (Tycho-2, Gaia, and 2MASS), to which we have devoted a strong effort until being satisfied 
\cite{Maiz05b,Maiz06a,Maiz07a,Maiz17a,MaizPant18,MaizWeil18}. The right panel of Fig.~\ref{fig4} shows how well this works even with a preliminary version in which
we used (a) aperture instead of PSF photometry and (b) the $\sim 100$ object calibration sample and only six filters (it was done before Gaia DR2 so no 
$G_{\rm BP}$+$G_{\rm RP}$ photometry was available and we used the calibrations of \cite{Maiz07a} and \cite{Maiz17a} instead of those of \cite{MaizPant18} and 
\cite{MaizWeil18}).

We also use several secondary calibration mechanisms:

\begin{itemize}
 \item Comparison of results for the same field using different exposure times and air masses to check for anomalous detector effects or atmospheric extinction.
 \item Comparison between adjacent fields. The GALANTE strategy leaves a generous overlap ($\sim 12^\prime$) between fields to allow for the identification of 
       possible zero point offsets.
 \item Use of spectrophotometric standards. Note that the GALANTE FOV is large enough that $\sim 10\%$ of the fields have standards of one type or
       another. Also note that we have recently increased the sample of spectrophotometric standards \cite{MaizWeil18}.
 \item Use as CHORIZOS input the temperature and gravity of objects with accurate spectroscopic classifications, especially early-type stars for which the 
       intrinsic SED is well known. This leads to reduced uncertainties for the predicted magnitudes. For this purpose we use the spectral classifications from
       the Galactic O-Star Spectroscopic Survey (GOSSS, \cite{Maizetal11,Maizetal16,Sotaetal11a,Sotaetal14}).
\end{itemize}

\section{Objectives and planning}

$\,\!$\indent The main objective of GALANTE is to identify all Galactic O+B+WR stars down to magnitude 17 and estimate their $T_{\rm eff}$. We will cross-match all
OBA stars with 2MASS and measure their $E(4405-5495)$ and $R_{5495}$. We will coordinate our efforts with the Stellar, Circumstellar, and Interstellar Physics
WEAVE survey and with GOSSS to acquire follow-up spectroscopy of the newly found O+B+WR stars.

Some additional objectives include (a) a magnitude-limited catalog of emission-line stars, (b) the IMF of large-area clusters and associations, (c) a 
continuum-subtracted H$\alpha$ map with subarcsecond pixels, and (d) cross-calibration with Gaia.

GALANTE NORTE started taking data in 2016 and GALANTE SUR in 2018. If weather behaves, we should complete the northern survey in 2021 and the
southern one in 2023. For the long-term future several extensions are possible: deep surveys of interesting regions, multiple epochs, and additional filters are
some of the possibilities.

% Do not delete the next line
\small  % Do not delete
%
%%% Comment the following line if you do not have acknowledgments.
\section*{Acknowledgments}   % Do not delete if you declare acknowledgments
%
%%% ACKNOWLEDGMENTS
%%% ACKNOWLEDGMENTS
Based on observations made with the JAST/T80 telescope at the Observatorio Astrof{\'\i}sico de Javalambre, in Teruel, owned, managed and operated by the Centro 
de Estudios de F{\'\i}sica del Cosmos de Arag\'on.
J.M.A., E.J.A., and A.L. acknowledge support from the Spanish Government Ministerio de Ciencia, Innovaci\'on y Universidades through grant
AYA2016-75\,931-C2-1/2-P. 
R.H.B. acknowledges support from DIDULS project PR18143.
%R.H.B. acknowledges support from the ESAC Faculty Council Visitor Program. 
%
% Do not delete the next few lines

%
\end{document}